\documentclass[runningheads]{llncs}
\usepackage[T1]{fontenc}
\usepackage{graphicx}
\usepackage{amsmath}
\usepackage{mathabx}
\usepackage{hyperref}
\usepackage{fancyhdr}
\fancypagestyle{firstpage}
{
    \fancyhf{} 
    \fancyfoot[L]{Pre-print of paper presented at \textit{\href{http://rkde2023.isti.cnr.it}{RKDE Workshop} @ ECML PKDD 2023 \\ (proceedings forthcoming)}}
    \fancyhead[L]{}
    \fancyhead[R]{}
      
}

\begin{document}

\title{PICA: A Data-driven Synthesis of Peer Instruction and Continuous Assessment}
\titlerunning{PICA: A Data-driven Synthesis of Peer Instruction and Cont.\ Assessment}

\author{Steve Geinitz\inst{1}\orcidID{0000-0002-4327-8737}}
\authorrunning{S. Geinitz}
\institute{Metropolitan State University of Denver, Denver CO, 80204, USA \\
\email{geinitz@msudenver.edu}}

\maketitle              

\thispagestyle{firstpage}

\begin{abstract}
Peer Instruction (PI) and Continuous Assessment(CA) are two distinct
educational techniques with extensive research demonstrating their
effectiveness. The work herein combines PI and CA in a deliberate and novel
manner to pair students together for a PI session in which they collaborate on a CA
task. The data used to inform the pairing method is restricted to the most
previous CA task students completed independently.  The motivation for this
data-driven collaborative learning is to improve student learning,
communication, and engagement. Quantitative results from an investigation of the
method show improved assessment scores on the PI CA tasks, although evidence of
a positive effect on subsequent individual CA tasks was not statistically
significant as anticipated.  However, student perceptions were positive,
engagement was high, and students interacted with a broader set of peers than
is typical.  These qualitative observations, together with extant research on
the general benefits of improving student engagement and communication (e.g.
improved sense of belonging, increased social capital, etc.), render the method
worthy for further research into building and evaluating small student learning
communities using student assessment data. 

\end{abstract}

\section{Introduction}
The ultimate goal of a teaching professional is to have students effectively
learn the subject content at hand. Peer Instruction (PI)~\cite{mazur97} and
Continuous Assessment (CA)~\cite{butler10} are two distinct pedagogical
techniques shown to facilitate this goal. The work described herein is a
deliberate, data-driven synthesis of PI and CA that places students into pairs
or groups to complete a (collaborative) CA task. 
The grouping of students is
determined by a short-term profile of each student's understanding on an
earlier (independent) CA task they have completed.  
The name, \textit{PICA}, derives from the fact that a PI session is centered
around one particular CA task, which together is referred to as a,
PICA session. 
This method has been tested
in a lower-division Discrete Mathematics course in an undergraduate Computer
Science degree program at a large institution serving a diverse student population.
The overall goal was to improve student learning, while also attempting to foster
student engagement and communication. 

\subsection{Peer Instruction}
Although PI originated in higher education and in the domain of
Physics~\cite{crouch01,mazur97}, its effectiveness has been demonstrated for
primary and secondary education~\cite{schell18}, as well as for a broader field
of STEM subjects~\cite{vickrey15}.  PI is typically implemented in the
classroom by: (1) the instructor presenting a multiple-choice problem over a
specific concept, (2) students independently answering, typically with a
clicker-style tool, (3) students pairing up and discussing their answers and
rationale, (4) students answering the question again~\cite{crouch07}.  A more
general technique also shown to be effective is known as Active Learning
(AL)~\cite{bonwell91,brown14}.  Much of the research on PI and AL has focused
on the aggregate effects. However, recent work has highlighted heterogeneous
effects that suggest university students from underrepresented backgrounds
(e.g. first-generation students, racial or socioeconomic minority students,
etc.) may benefit even more from PI/AL than a traditional university
student~\cite{theobald20}. The underlying causes of these heterogeneous effects
may be related to some of the benefits cited by long-standing research on more
general peer-to-peer education. Goldschmid and Goldschmid~\cite{goldschmid76}
describe the improved socialization that can be gained from student-led
instruction. It is also known that students from historically underrepresented
groups begin university with less social capital and a lower sense of belonging
than their peers~\cite{seider15,walton11}, and that these qualities are related
to students' intrinsic motivation and self-efficacy~\cite{freeman07}.
Regardless of the subpopulation to which a student belongs, there exists
evidence that students with lower metacognition can benefit even more from peer
discussion~\cite{molin20}. 

\subsection{Continuous Assessment}
Assessment tools in education can be designed and used in various
ways~\cite{brown12}, although there is little contention on their
utility~\cite{hernandez12}.  Despite the varied perspectives, the greatest
benefits seem to be gained when assessments are frequent, low-stakes, and the
concepts are spaced and interleaved~\cite{butler10,roediger11}, which are the
characteristics we have adopted when using CA.  Another characteristic of CA we
have embraced is that students' retrieval of information should involve effort,
as this has been seen to maximize learning~\cite{brown14}.  Many of these
characteristics have successfully been adopted in modern settings as well, such
as Mobile Learning~\cite{hwang11}. In general, the ease
with which an instructor can employ such assessment tools has only increased with
the use of Learning Management Systems (LMS)~\cite{ghani19}, which has opened 
the door to explore new implementations of CA.

\subsection{Collaborative Exams}
Collaborative Exams (CE) are a form of assessment that address some of the
aspects of PI discussed above by having students work jointly on a course
examination. How CE is utilized can vary with some implementations allowing
students to collaborate on only certain sections of an exam with one
attempt~\cite{eaton09}. The most common approach, however, is akin to PI in
that students first take an exam independently, then take the same, or a
similar, exam a second time with one or more peers~\cite{bloom09,rieger14}.  
One potential disadvantage of CE is that examinations are heavily weighted with
respect to students' final grades, which can negatively impact their attitudes
and perceptions towards this form of collaboration~\cite{scherman14}.  It is
also in direct opposition to the low-stakes quality of CA since students'
anxiety can interfere with their test-taking abilities~\cite{duraku17} and
inhibit meaningful discussion from taking place (since the high-stakes format
places greater importance on simply getting a correct solution).  Formal
accountability mechanisms have been used with some success in improving student
interaction~\cite{chou15}, but this has been in more of a PI setting where the
stakes for students are relatively lower than with exams, and the frequency is
higher.

Student anxiety is one aspect that must be considered as it has been
increasingly prevalent in recent years~\cite{duraku17} and has accelerated with
COVID-19~\cite{li21}.  Although evidence seems to suggest students from
minority ethnic/gender groups do not suffer from anger/anxiety/depression at
higher rates~\cite{rosenthal00,zeidner90}, or that these groups are able to
deal with such psychological symptoms more readily~\cite{hayes11}, the
potential of increasing student stressors remains a concern for us when
considering how to utilize any form of collaborative assessment. 

\subsection{PICA Overview}
PICA sessions are intended to maintain the quality and frequency of
student-to-student interaction achieved with PI, while incorporating the
formality of CA with a low-stakes nature to mitigate potential student anxiety.
Thus, the frequency, duration, and structure of a PICA session already
distinguish this method from previous research in the intersection of PI and
CA~\cite{cao17a,chou15}.  This study goes one step further by investigating
whether student pairing can be done in an informed manner  to maximize
learning. The intuition behind the method used to assign student pairs is based
on complementary knowledge, or diversity of thought. In a hypothetically ideal
pairing assignment, one student would have scored three points out of five on
an independent quiz in a dyad by answering the first three questions correctly,
while the second student would have scored two points by answering only the
final two questions correctly. Thus, it is not their overall score, but rather
their five-dimensional score vector that is used with an aim of creating pairs
of students possessing complementary knowledge. In this example, the two scores
vectors are orthogonal yet have comparable magnitudes.  This is to help
mitigate the risk of pairing two students with disparate abilities. Earlier
work differs on whether collaborative assessments benefit low, mid, or high
performing students equally~\cite{cao17a,dahlstrom12}, but agrees that
effectiveness varies with the relative abilities of the students. Work in other
domains has also shown pairing a low performer with a high performer can cause
issues~\cite{lopez22}. Hence, why the example with orthogonal vectors but
similar magnitudes is seen as ideal. While other student data could be used to
inform the pairing, such as grade history, demographic/socio-economic information, etc., this method takes a more
data-sensitive approach by utilizing only the most recent quiz scores. 

Latuliple et al.\ attempted something similar by using PI, collaborative
quizzes, and other student-team activities and gamification~\cite{latulipe15}, and 
yielded positive outcomes for first-year undergraduate students.
However, with the teams randomly generated only once at the start of the
semester, the students' knowledge and capabilities did not inform the grouping.
Additionally, students were not presented with opportunities and/or incentives
to interact with others outside of their own team.

It is worth noting that this research has been both motivated and enhanced by
the recent health epidemic. First, the PICA sessions were introduced when
returning to face-to-face classes as a way to foster human-centered
communication and to (re)grow students' social networks, since the pairing
method rarely yields the same student pairs. Second, an artifact of the year of
remote learning has been that we, and many instructors at our institution, have
continued to offer students the option to attend class lectures virtually.
While students are formally encouraged to attend in person, some choose to
attend virtually from time to time.  Students attending class virtually on the day of a
PICA session are still given the opportunity to take the collaborative CA, although they
are not assigned a partner and therefore complete it without the benefit of
face-to-face collaboration with a peer. The combination of in-person pairs and
remote individuals has provided a type of treatment vs. non-treatment
comparison for the PICA sessions. Unfortunately, due to practical limitations and ethical
concerns, a true randomized test and control comparison was not possible.  Thus,
there are two general research questions sought to be answered regarding the
method. \\

\noindent \textbf{RQ1:} Do students taking part in a PICA session generally see
greater increases in learning than students that do not? \\

\noindent \textbf{RQ2:} Is complementary knowledge positively correlated with
the size of the learning increase? \\


The rest of the paper is organized as follows. Section 2 describes the PICA method and
its implementation in a Discrete Mathematics course. Section 3 reports the
results of the study. Section 4 concludes by exploring potential extensions of this work.


\section{Methodology}

\subsection{PICA Method}

Frequent and structured assessments are at the core of this method and are in
the form of quizzes. Quizzes come in sets of two, with one set referred to as a quiz dyad. 
One dyad is
administered nearly every week over a semester course. In our course, this
yielded ten quiz dyads, which were seen by students as Quiz 1a, Quiz 1b, Quiz
2a, \dots , Quiz 10a, Quiz 10b.  The '\textit{a}' and '\textit{b}' refer to the
first and second quiz in a dyad, respectively.  Students complete a-quizzes independently and b-quizzes collaboratively. 
Each quiz is comprised of
questions focusing on a single concept or topic. A quiz dyad may have one or
more pairs of isomorphic questions~\cite{porter11} across the a-quiz and
b-quiz, i.e., questions that are distinct but attempt to interrogate a
student's knowledge of the same specific topic or concept (Figure 1).

\begin{figure}[h]
  \centering
  \includegraphics[scale=0.28]{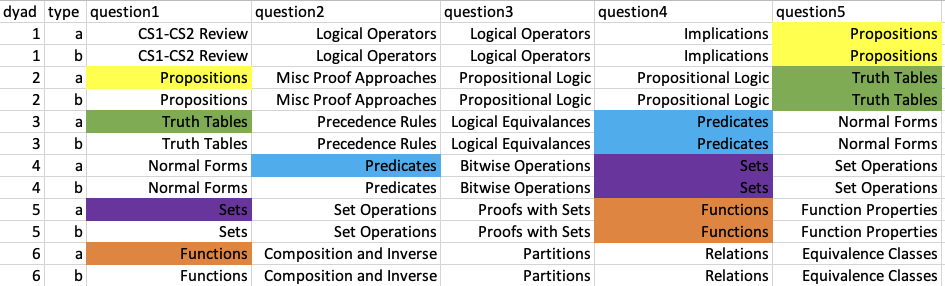}
  \caption{Six quiz dyads with the specific concept covered by each quiz
  question. An 'a' quiz is taken by students independently outside of class
  time. The subsequent 'b' quiz covers the same concepts but is taken by
  students collaborating in pairs during class time.  Coloring indicates questions 
  deliberately crafted to test students on the same concept or topic in order to later 
  quantify the effects of collaboration.}
  \label{fig:dyads}
\end{figure}

\begin{figure}[h]
  \centering
  \includegraphics[scale=0.34]{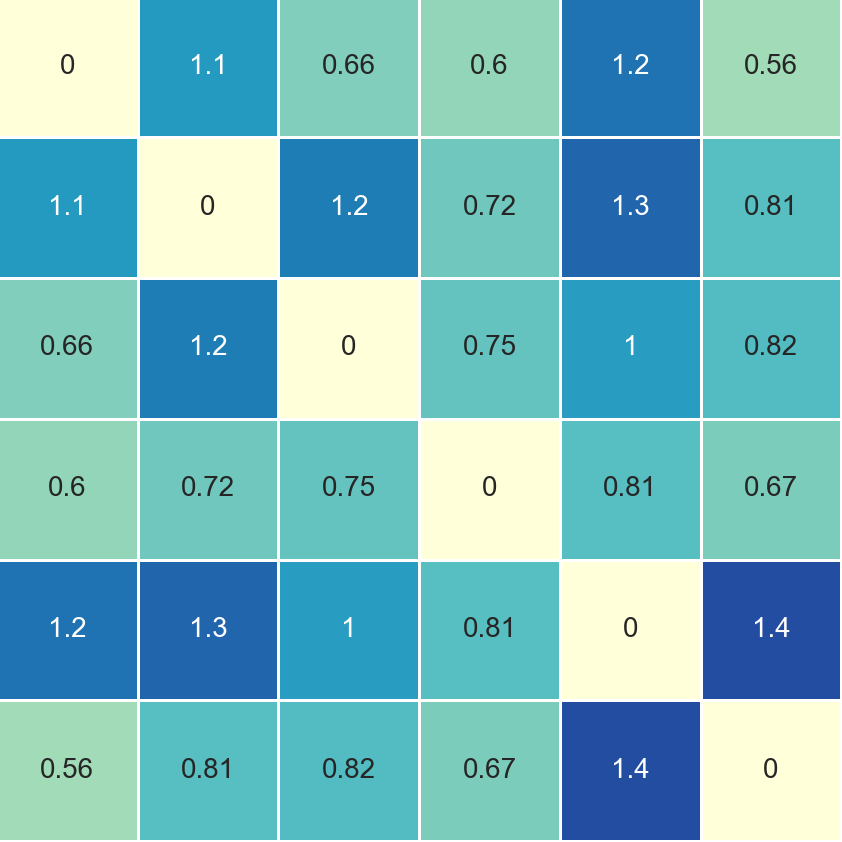}
  \caption{A sample distance matrix of the $n(n-1)/2$ pairwise distances
  between the five-dimensional quiz score vectors for a subset of six students 
  (from $n=34$ students in total).  Each row and column
  corresponds to a student  (identifiers have been excluded). The number and
  color in a cell represent the Euclidean distance between score vectors.  }
  \label{fig:distmat}
\end{figure}

Again, the first quiz in a dyad is completed by students individually outside of class
time. During the subsequent class meeting, this first quiz is reviewed and
discussed as a class, and students then take the second quiz in the dyad.
Before this second quiz is administered, all students physically present in the
classroom are deterministically assigned a partner to collaborate with.
Students are given roughly twice as much time to complete the second quiz in a
dyad to facilitate in-depth discussion with their partner. The assignment of
partners is intended to pair students whose knowledge or performance on the
first quiz was complementary. To accomplish this, the students'
five-dimensional score vectors from the first quiz are used to compute the
Euclidean distance between all possible pairings of students (Figure
\ref{fig:distmat}).  These $n(n-1)/2$ possible pairs are then searched to
create $\lfloor \frac{n}{2} \rfloor$ pairs and, possibly, one group of three
students.  Various algorithms/methods could be employed, although for this
study a practical approach was used to create complementary student pairs while
maintaining a small amount of variability (in order to study the effect of distance while
also preventing students with disparate overall a-quiz scores from being paired
together). The specific method employed begins by taking the maximum distance
(from the set of $n-1$ distances) for each of the $n$ students. These $n$
distances represent $n$ pairings, and each student is in at least one of the
pairings. The median distance among these $n$ is selected, and the two
corresponding students are paired with one another.  These two students are
then removed and the same steps are repeated with the remaining $(n-2)(n-3)/2$
distances. This method has been implemented in an associated software 
package~\cite{picata22} to allow for quick and efficient pairing of students on  
the day of the b-quiz. 

One final yet important aspect of this method is how the instructor presents a
collaborative b-quiz to students.  Once students have been told who they are
partnering with, but before they stand up to move to sit next to their partner,
students are reminded of the motivation for this method. Namely, that the
pairing is based on complementary a-quiz scores (though not too disparate),
that it is both beneficial for them to hear and to explain a concept in their
own words, that the practice of explaining these concepts in their own words
helps them to establish better technical communication skills, and that
communicating with someone new can be beneficial.  Small/brief words of
encouragement are also given (e.g. emphasizing a growth mindset). These seemingly extraneous factors 
have recently been shown to maximize the positive
effects Active Learning approaches~\cite{theobald20}. 

\subsection{Implementation} 
\begin{figure}[h]
  \centering
  \includegraphics[scale=0.21]{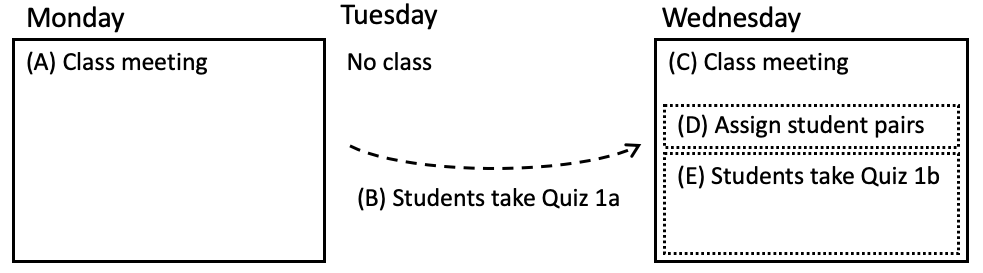}
  \caption{Weekly schedule of the course in which the PICA method was tested and the
  first quiz dyad was administered.  (A) and (C) represent class meetings while
  (B) represents the roughly 48-hour period in which Quiz 1a is available to
  students. (D) is when the instructor executes code to download Quiz 1a results
  and generate student pairings (for students present in classroom).  (E) is the
  25-30 minute period in which students in class collaborate on Quiz 1b, and
  remote students take it independently. }
  \label{fig:schedule}
\end{figure}

The PICA method has been tested in a Discrete Mathematics course with $n=34$
students over a 15-week semester (one potential limitation to the PICA sessions
is the relatively small class size and a room allowing students to easily move
around).  Ten quiz dyads were administered over this time. Students were given
10-12 minutes to complete each (individual) a-quiz and 25-30 minutes for each
(collaborative) b-quiz. Each quiz consisted of five questions, with each
question often having multiple parts (e.g.\ 2-4 multiple choice/matching/etc.).
Figure \ref{fig:schedule} illustrates the timeframes in which the two quizzes
in a dyad were administered in a typical week in a course meeting two days a
week. 

The concepts covered by each question on each quiz were carefully selected and
documented to ensure a) the score vectors could effectively span the five
dimensions, and b) the effects of collaboration for a individual could be
quantified later on using a future a-quiz. Figure 1 contains the concepts
covered by the first six quiz dyads and highlights some of the concepts
repeated across dyads.

The Canvas LMS \footnote{https://www.instructure.com/canvas} 
was used to administer all quizzes, including the PICA session
b-quizzes. Although students collaborated on these b-quizzes, they each input
answers into Canvas separately. The individual input ensured student
participation to at least some degree. But, an additional mechanism was also used
to encourage active discussion, which was to institute a bonus for
collaboration~\cite{chou15}. A one-point bonus (i.e. 20\% of the five points
possible) was awarded to each student in a pair if they had the same answers
across all five questions. Students were given this bonus regardless of
correctness so that, for example, a score of 5/5 points could still be achieved
even if students missed one question entirely but had all five answers
matching. This bonus mechanism was not employed in earlier pilot runs of this
study. However, instructors noticed a significant increase in the volume and depth
of student discussions when this bonus was used. 

The Canvas API Python module~\footnote{https://github.com/ucfopen/canvasapi}
has been utilized by the PICA software~\cite{picata22} to automate the
process of downloading a-quiz results, creating distance matrices (Figure 2),
assigning student pairs, and awarding b-quiz bonus points. The pairing
considered only those students physically present in the classroom on the day a
PICA session took place. This required as input a list of students present,
which was logged by the instructor before class and/or during a break given
just before the b-quiz was administered. This highlights a potential limitation
to this method being used in larger classes. A similar approach used by Chou
and Lin~\cite{chou15}, which employed a pairing method that relied on students
logging/registering their presence in the classroom may scale to larger classes
more readily.  Although their method required that students be in close
physical proximity to one another in order to be grouped together, which limits
the pairing combinations that can be allowed.

\section{Results}
The results of PICA as a teaching technique are evaluated according to the
research questions of Section 1. RQ1 is examined using the change  
between a student's a-quiz and b-quiz scores within a dyad. As stated previously,
a student attending a class meeting remotely would thus not have a b-quiz
partner but would be allowed to take the b-quiz remotely and independently.
These students are compared to those that were present in the classroom and
did have a partner to collaborate with. These two groups of students form a sort
of quasi treatment and control and are referred to as such. True test and control groups, meaning a randomized control trial,
were not possible due to practical and ethical concerns. 
The change between quiz
scores within a dyad are measured using the Modified Normalized Gain (MNG)~\cite{marx07}. 
Normalized Gain (NG) is intuitively stated as the
"\textit{amount learned}" divided by the "\textit{amount that could have been
learned}", and is formally defined by Equation (\ref{eq:ng}).  NG can lead to large negative values when
an a-quiz score is greater than the b-quiz, or be undefined for a perfect a-quiz
score. Hence, the the adoption of the MNG, which is stated by Equation 
(\ref{eq:mng}). 
\begin{equation}
\textrm{NG} = \frac{\textrm{Quiz B } - \textrm{Quiz A}}{5.0 - \textrm{Quiz A}}
\label{eq:ng}
\end{equation}
\begin{equation}
\textrm{MNG} = 
\begin{cases}
0 & \textrm{ if Quiz B } = \textrm{Quiz A } \\
\textrm{NG}  & \textrm{ if Quiz B } > \textrm{Quiz A} \\
\frac{\textrm{Quiz B } - \textrm{Quiz A}}{\textrm{Quiz A}} & \textrm{ if Quiz B } < \textrm{Quiz A }
\end{cases}
\label{eq:mng}
\end{equation}

\begin{figure}[h]
  \centering
  \includegraphics[scale=0.39]{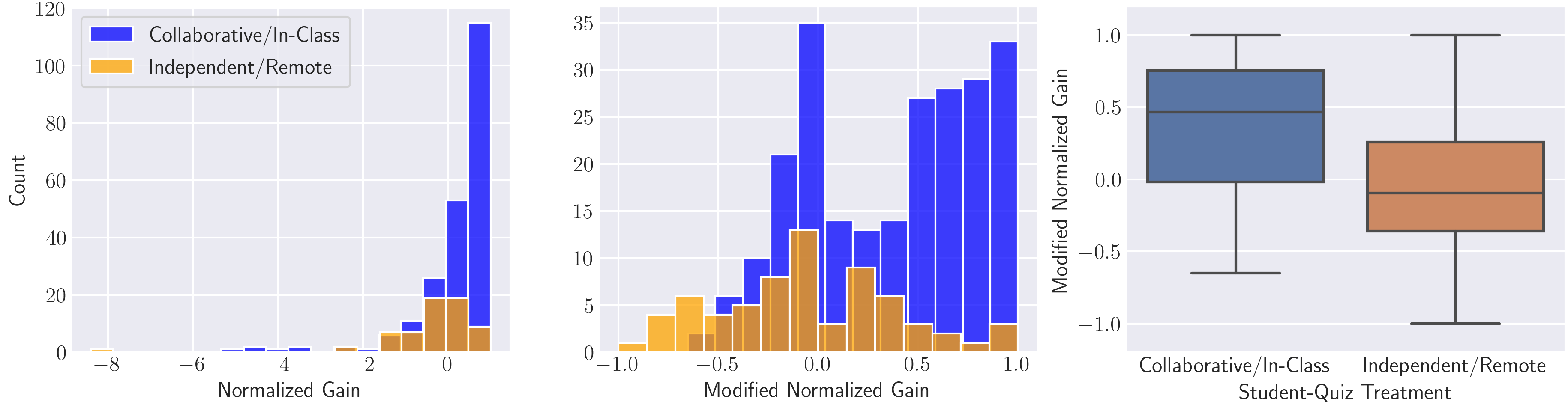}
  \caption{Normalized gain (left) and modified normalized gain (center), and a
  boxplot of MNG (right) for each quiz dyad and each student. Each plot
  represents students taking a b-quiz independently and remotely (orange,
  $n=68$) and students taking a b-quiz collaboratively in class (blue,
  $n=232$).}
  \label{fig:rq1a}
\end{figure}

\noindent
The NG and MNG values for the quasi treatment and control are presented in
Figure \ref{fig:rq1a}.  From the histograms and boxplot, it is clear there is a
difference in the distributions of NG and MNG for students taking a b-quiz
collaboratively versus those taking it independently. There was also a
statistically significant difference in MNG when using either a parametric test
to compare the means (Student's t-test: $\widebar{x}_{treat} = 0.37$ vs
$\widebar{x}_{cont}=-0.05$, $p = 9.1e^{-10}$) or a nonparametric test to
compare the distributions of the two (Mann-Whitney: $p = 9.4e^{-10}$).  This
result is unsurprising as there is likely an influence from
confounding factors.  Namely, students taking a b-quiz at home may lack
motivation, may not have resources to succeed, etc. It is important to note, however, that it was not always
the same set of students that attended remotely.  The number of unique students
represented in the treatment and control groups was $29$ and $16$,
respectively. However, there were also $12$ students who were present in both
groups (i.e. took at least one b-quiz in class and at least one remotely). It
is also important to note only completed quiz dyads were used, meaning that if
a student missed either an a-quiz or b-quiz, then their quiz dyad was excluded.
Furthermore, all quiz scores were strictly greater than zero, so there were no
cases in which an a-quiz or b-quiz was equal to zero. Finally, it is important
to note that all analyses in this study used b-quiz scores before the bonus
point was awarded. 

\begin{figure}[h]
  \centering
  \includegraphics[scale=0.39]{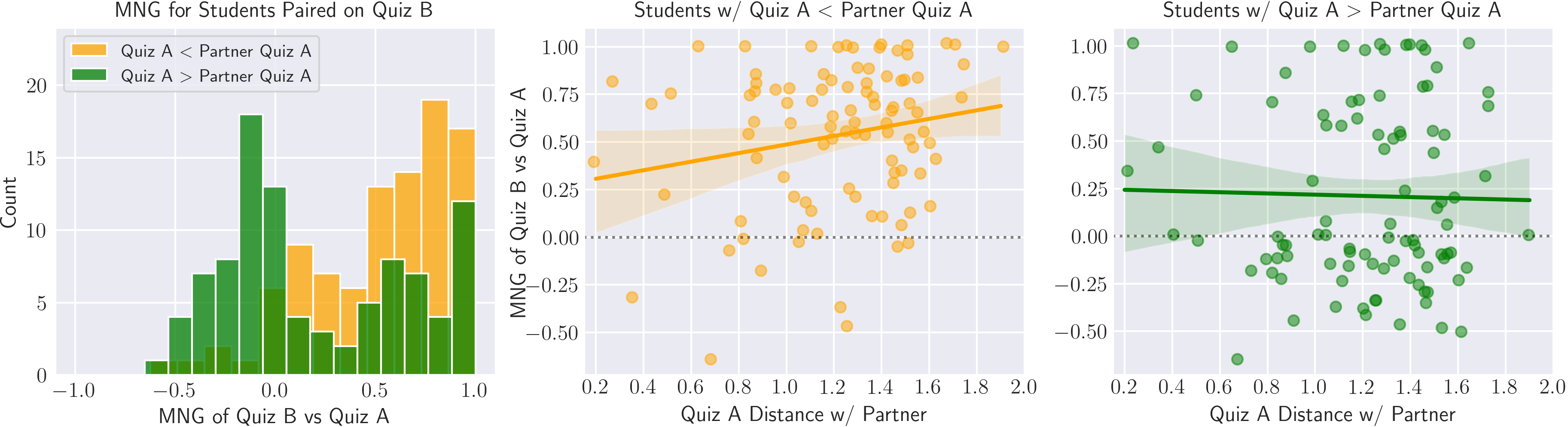}
  \caption{Modified normalized gain (MNG) split by students whose a-quiz score
  was lower (yellow)/higher (green) than their partner's (left). These same students'
  MNG is also plotted against the a-quiz distance to their partner (center,
  right) with the approximate confidence interval of the relationship shaded in
  the respective color.}
  \label{fig:rq2a}
\end{figure}

To address RQ2, which examines the relationship between a pair of students'
a-quiz distances and the improvement each sees when taking a b-quiz
collaboratively, it is helpful to first split students depending on whether
they had a higher or lower a-quiz score than their partner for the
corresponding b-quiz.  This is because it is sometimes the case that the
student who scored higher than their partner may have less opportunity to
improve.  While this is mitigated to some degree by using the MNG, it still
occurs when a b-quiz score was less than the corresponding a-quiz, and the
third case of Equation (\ref{eq:mng}) is applied. This is done by splitting the
treatment data of Figure \ref{fig:rq1a} (in blue) into two subsets, as seen in
Figure \ref{fig:rq2a}. Note that for simplicity, the four cases in which two
students paired together had the same score on an a-quiz were excluded.
Additionally, all data points representing cases in which three students were
paired together were also excluded. From the $232$ cases above, this leaves
$192$, or $96 = 192/2$ cases in which a student had an a-quiz lower than their
partner and $96$ in which it was greater. 

\noindent
To answer RQ2, the linear relationship between MNG and distance may be used and
tested for statistical significance.  Figure \ref{fig:rq2a} presents the
approximate confidence intervals of the slope of this relationship for students
whose a-quiz score was lower than their partner's a-quiz score (center, in
yellow) and those for which it was higher (right, in green). The t-test to see
whether this slope was significantly different from zero yielded a p-value of
$p = 0.045$ in the case of the former (marginally significant), and $p = 0.822$
in the latter (not significant).  These results are, again, not surprising in
that when there is a greater distance between two students' a-quiz scores, then
one student will have inevitably scored lower than their partner, and will thus
have more opportunity to improve. 

\begin{figure}[h]
  \centering
  \includegraphics[scale=0.39]{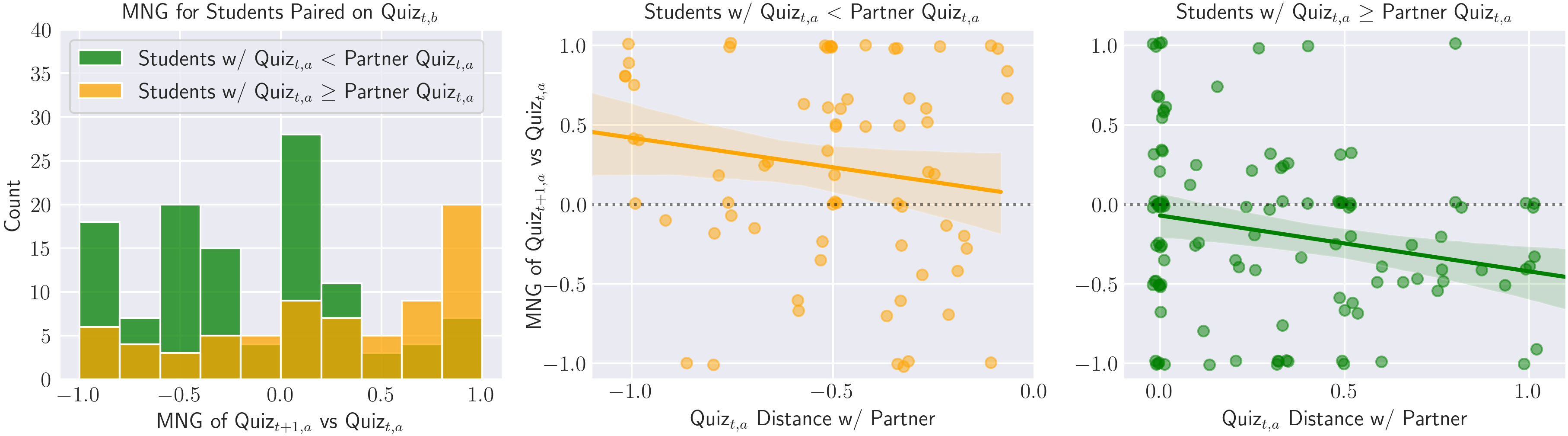}
  \caption{Modified normalized gain (MNG) for each student for individual isomorphic questions. This is split by students whose a-quiz score 
  for the given isomorphic question was lower (yellow)/higher (green) than their partner's (left). 
   MNG is also plotted against the absolute distance to their partner on the given question (center, right). Although every attempt was made to create
   isomorphic questions, the negative trend (center, right) may be due to students' perceived increase in question difficulty.}
  \label{fig:rq2b}
\end{figure}

It is also possible to analyze the quiz results to understand whether a PICA
session helped a student to improve their individual understanding. To
accomplish this we calculate the gain for a specific concept across two dyads
for which isomorphic questions were created.  Figure \ref{fig:dyads} highlights
several concepts where this was possible. For example, a student's
understanding of the concept of \textit{Truth Tables}, could be assessed with
the gain between Quiz 2A Question 5 and Quiz 3A Question 1 (with the Quiz 2B
PICA session also covering this concept). 
The notation is modified slightly to illustrate this more generally.  Namely,
this means calculating the MNG for individual questions that cover the same
concept on Quiz$_{t,a}$ and Quiz$_{t+1,a}$, when a student was engaged in a
PICA session on Quiz$_{t,b}$.  This was done for several concepts in which
isomorphic questions were created across quizzes and the results combined to
obtain a reasonable sample size of data points (Figure \ref{fig:rq2b}). There
were eight concepts for which isomorphic questions were created across quiz
dyads. The Discrete Mathematics concepts were: \textit{Propositions},
\textit{Truth Tables}, \textit{Predicates}, \textit{Sets}, \textit{Functions},
\textit{Summations}, \textit{Algorithm Analysis}, and \textit{Asymptotic
Notations}.  An example of isomorphic questions for \textit{Summations} starts with Quiz
7a, in which students were asked to identify the correct value each of the following three
summations, $S_1$, $S_2$, and $S_3$,  would evaluate to using a Canvas matching
question\footnote{https://community.canvaslms.com/t5/Instructor-Guide/How-do-I-create-a-Matching-quiz-question/ta-p/918}.
\begin{displaymath}
\textrm{Quiz 7a, Question 4:} \qquad S_1 = \sum_{k=1}^{8} k \qquad
S_2 = \sum_{k=1}^{8} 4 \qquad
S_3 = \sum_{k=0}^{8} 2^k 
\end{displaymath}
Quiz 8a then presented students with a question in which they were asked to calculate a
summation on their own and input the resulting value. 
\begin{displaymath}
\textrm{Quiz 8a, Question 2:} \quad \quad \quad \quad \quad S = \sum_{k=1}^{a} k- \sum_{k=1}^{b} k \qquad \qquad \qquad \quad
\end{displaymath}
For this type of formula
question\footnote{https://community.canvaslms.com/t5/Instructor-Guide/How-do-I-create-a-Simple-Formula-quiz-question/ta-p/1233}
in Canvas, unique values for $a$ and $b$ are randomly generated (with $a > b$)
for each student that takes the quiz.  To address RQ2 the distance between
students is again calculated, although now it is based on a single question.
Since when looking at a single question there is a high probability both
students earned the same points, a qualifier of greater than or equal is now
used.  Also note that the distance is now calculated based on the simple
difference for that single question from Quiz$_{t,a}$ so that students scoring
less than their partner will have a distance in the interval $[-1,0)$ while
those scoring the same or more than their partner will have a distance in the
interval $[0,1]$. The number of data points in these intervals were 73 and 117,
respectively (Figure \ref{fig:rq2b}). 

For students scoring the same or higher than their partner on the a-quiz
question (Figure \ref{fig:rq2b}, right), the results are roughly consistent
with those presented in Figure \ref{fig:rq2a} (right).  However, for students
scoring less than their partner on the a-quiz question, the results now suggest
no relationship exists between distance and MNG (Figure \ref{fig:rq2b},
center).  In other words, the PICA session does not seem to improve a student's
understanding over the concept on a future assessment.  One possible reason for
not seeing greater gains was perhaps that quiz questions over the same concept
gradually increased in difficulty over time, as seen with the quiz questions on
summations.  As instructors, we consider these questions to be isomorphic, but
also acknowledge that students might perceive the second to be more difficult.
This highlights a challenge observed by others  in that, "\textit{questions
deemed isomorphic by instructors are not necessarily isomorphic to
students}"~\cite{porter11}. This is a hurdle that must be addressed in
future investigations of this method. 

\section{Conclusion}

Despite the quantitative results of the PICA method, we are encouraged by 
the overall findings and eager to continue. The anecdotal evidence of increased 
student engagement is, on its own, sufficient to warrant further research. 
Thus, one modification to be instituted in the next iteration of this work
will be to request qualitative feedback from students after each PICA session.
This would provide direct student perspectives on the pairing method and any
effects.  Akram et. al.~\cite{akram22} have had students complete
self-assessments after larger projects and instructors provided feedback to
help increase students' sense of belonging, professional confidence, etc.
While the frequency of PICA sessions is too high to do this, the qualitative
feedback from students alone could be useful to assess whether some of the
difficult-to-quantify benefits of the PICA method do exist. A specific research
question of interest that might be answered with such feedback pertains to the
possible psychological effects on students of knowing they are deliberately
paired with someone possessing complementary knowledge.  This could also 
yield comparisons with other types of pairing methods, such as students having 
greater control over the pairing method via explicit social
comparison~\cite{akhuseyinoglu22}.

Another extension of this work will be to increase the signal on student
profiles to improve the pairing method. While it may be interesting to compare
the current pairing method with one that uses more in-depth student profiles
(e.g. grade history, demographics, socio-economic status, etc.), we are
reluctant to take this approach due to data sensitivity. A different approach
to producing richer student profiles could instead be done by collating
additional information via a concept-inventory~\cite{taylor20} type of
pre-assessment at the start of a semester and/or utilizing all prior CA results
to uncover temporal patterns in a student's learning progression~\cite{rotelli22}.

There are also various opportunities to utilize large language
models (LLMs). We have already experimented with an LLM to create vector representations of 
the quiz questions and cluster these in an attempt to automate some of
the manual work of documenting the concepts or topics covered (Figure \ref{fig:dyads}). 
However, identifying isomorphic questions will likely still require a subject matter expert. 
Another possible application of LLMs would be
to generate personalized follow-up questions after an assessment.
Longer term, we hope to extend this further to mediate an online
discussion between the pair of students over a messaging platform, which could help to 
foster student-to-student learning and the development of student learning communities.

\section*{Acknowledgements}

We thank our institution for funding an early pilot version of this study,
as well as our student, Nicholas Ryan, who performed some initial data exploration during that time and was a contributor to the PICA software.
We also thank our Human Subjects Research Projects office for their support in
obtaining IRB approval for this study. All de-identified data, code, and
materials are available upon request. 


\bibliographystyle{splncs04}

\bibliography{pica}

\end{document}